\documentclass[%
 aip,
 amsmath,amssymb,
 reprint,%
]{revtex4-1}
 \usepackage{color}
\usepackage{graphicx}% Include figure files
\usepackage{dcolumn}% Align table columns on decimal point
\usepackage{bm}% bold math
\usepackage[utf8]{inputenc}
\usepackage[T1]{fontenc}
\usepackage{mathptmx}
\usepackage{etoolbox}

\makeatletter
\def\@email#1#2{%
 \endgroup
 \patchcmd{\titleblock@produce}
  {\frontmatter@RRAPformat}
  {\frontmatter@RRAPformat{\produce@RRAP{*#1\href{mailto:#2}{#2}}}\frontmatter@RRAPformat}
  {}{}
}%
\makeatother
\begin{document}

\preprint{AIP/123-QED}

\title{High-Precision Method for Characterizing Degree of Collimation and Beam Quality for Application in Cold Atom Gravimeter System}

\author{Nawaz Sarif Mallick}
\author{Anju}
%\author{Harish Rathore}
%\author{Abhishek Bhardwaj}
\author{Aishik Acharya}
\email{aishik.acharya@tcgcrest.org}
\affiliation{Centre for Quantum Engineering, Research and Education (CQuERE), TCG CREST, Kolkata-700091, India.}

\date{\today}% It is always \today, today,

\begin{abstract}
Highly collimated laser beam with excellent spatial quality is essential 
for quantum sensing experiments, where even small residual beam 
divergence can accumulate over distance and introduce significant systematic errors. 
In this article, we present the design and detailed characterization of a high
precision laser beam collimator developed for a cold-atom Gravimeter system, capable 
of producing an expanded laser beam with a diameter of 16 mm while achieving 
microradian level collimation accuracy through a five-degree-of-freedom (5-DOF) adjustment mechanism. 
The beam quality is evaluated using an ISO11146 compliant beam propagation measurement 
combined with Gaussian beam analysis to extract key parameters, including the beam waist, 
divergence angle, Rayleigh length, and beam quality factor $M^{2}$. 
The measured divergence angles of 0.006° (105 micro-radian) along the $x$ axis and 0.007° (122 micro-radian) along the $y$ axis confirm stable and well controlled collimation over long propagation distances. 
The demonstrated collimation architecture and characterization methodology provide a 
robust and scalable solution for cold-atom Gravimetry and other precision optical applications 
that require stable, high quality laser beams maintained over extended distances. 
\end{abstract}

\maketitle

%********************************************************************************************
\section{\label{sec:level1}Introduction}
%********************************************************************************************
Cold-atom Gravimeter represents one of the most demanding platform in quantum sensing, 
requiring exceptionally precise control over laser beam properties to achieve high sensitivity and measurement accuracy.
In such high sensitivity systems, highly collimated laser beam with excellent spatial quality are a fundamental requirement,
as they interact with atomic ensembles over long free fall distances and directly determine 
the fidelity of atom-light coupling.
Even small residual beam divergence or imperfect collimation can introduce 
undesirable variations in wavefront curvature, optical phase, and intensity distribution 
along the propagation direction. These imperfections accumulate over extended propagation distances, 
degrading the interferometric contrast and ultimately limiting the achievable measurement precision 
and sensitivity of the Gravimeter.
Therefore, achieving and maintaining accurate beam collimation, together with near diffraction limited beam quality, 
is essential for reliable operation of gravimetry system.

A key performance factor in our cold-atom Gravimeter is the efficiency of atom cooling and trapping. 
The number of atoms captured in the $^{87}$Rb Magneto-Optical Trap (MOT) is highly sensitive to 
the collimation quality of the cooling laser beams. Any deviation from ideal collimation can lead 
to reduced beam overlap, intensity inhomogeneity, and non-uniform radiation pressure, thereby 
degrading the cooling and trapping efficiency. 
This issue is particularly critical in cold-atom Gravimeter, where a large atomic 
population (typically on the order of $\sim 10^{8}$ atoms) is required to ensure optimal performance. 
A reduction in the number of trapped atoms directly lowers the signal-to-noise ratio of the atom 
interferometer and consequently limits the achievable measurement sensitivity and precision of the Gravimeter.

To address these challenges, we demonstrate the design, implementation, and detailed characterization 
of a high-precision laser beam collimator developed specifically for a cold-atom Gravimeter system \cite{Anju_2025}. 
We further present a robust experimental methodology for measuring beam parameters, including beam waist, 
divergence angle, Rayleigh length, and beam quality factor $M^{2}$.
This combined approach enables precise and repeatable control of beam propagation, which 
is essential for state-of-the-art quantum sensing experiments.

\begin{figure}[t]
\includegraphics[width=0.49\textwidth]{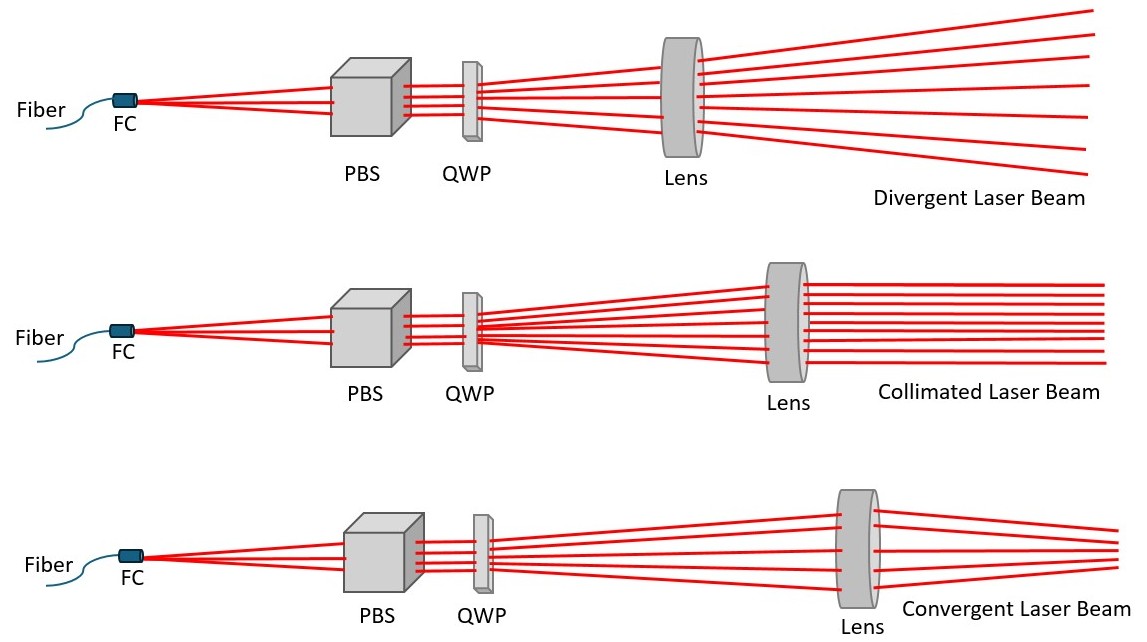}
\caption{\label{fig:epsart} Schematic representation of collimated and uncollimated laser beams.
FC: Fiber Coupler, PBS: Polarizing Beam Splitter, QWP: Quarter Wave Plate. }
\end{figure}

To better understand beam collimation procedure and its practical realization, we examine the 
fundamentals of Gaussian beam and laser beam propagation.
A laser beam appears collimated, divergent, or convergent because it is fundamentally 
a Gaussian optical wave governed by diffraction. Since the beam originates from a finite 
waist inside the laser cavity, it naturally spreads as it propagates, producing divergence. 
By appropriately positioning the lens and adjusting the focal plane along the optical axis, the 
diverging beam can be transformed either into a collimated beam or into a focused convergent beam, as 
illustrated figure \ref{fig:epsart}.
Note that the effective beam source must be positioned precisely at the focal plane of the collimating lens 
to obtain a collimated beam \cite{photonics13020142,Prakash10032006}.
However, achieving an exceptionally well-collimated laser beam over long distances requires 
high-precision alignment in the transverse ($x$ and $y$) and longitudinal ($z$) directions, along 
with precise control of the beam angle and tilt.
For this purpose, we designed and assembled a beam collimator using off-the-shelf components 
that enables high-precision alignment and produces an exceptionally well-collimated laser beam with a diameter of $16~mm$.
The resulting $16~mm$ diameter collimated laser beam is employed for the laser cooling and trapping 
of $^{87}$Rb atoms in our cold-atom gravimetry system.
However, the beam collimator is versatile and can be employed in a wide range of atomic experiments, as 
it is capable of producing collimated beams with varying diameters. 
The primary objective is to collimate the expanded beam with high precision, achieving 
microradian level angular accuracy and maintaining collimation over long propagation distances.
Apart from its application in cold-atom gravimetry, long-range beam collimation is 
also essential for several other high-precision instruments, including atomic fountain 
interferometer \cite{atoms9030058}, atom interferometric 
gyroscope \cite{10.10635.0198240, Gustavson_2000, PhysRevApplied.23.044001}
and Bloch-oscillation based atom interferometer \cite{PhysRevResearch.6.L032028} $etc$.

The article organized as follows.
In section \ref{sec2}, we present the detailed design and implementation of the beam collimator. 
The collimator incorporates high-resolution kinematic controls along with a finely adjustable 
achromatic doublet lens, enabling independent optimization of the beam position, propagation angle, 
and longitudinal focus. The resulting system produces a highly collimated laser beam with 
a divergence angle on the order of $100~\mu\mathrm{rad}$, demonstrating excellent long-range collimation performance.
In section \ref{sec3}, \ref{sec4}, and \ref{sec5}, we describe a robust experimental methodology 
for characterizing the beam parameters, including the beam waist, divergence angle, Rayleigh length, 
and beam quality factor. By combining Gaussian beam theory with an ISO 11146-based beam propagation analysis, 
we accurately determine the beam waist, divergence angle, and beam quality factor, providing 
a comprehensive assessment of the collimator performance.
In section \ref{sec6}, we present the implementation of the designed beam collimator 
in the cold-atom Gravimeter system.
Finally, the article concludes with a summary and final remarks in section \ref{sec7}.

%********************************************************************************************
\section{\label{sec2} Details of high-precision Beam Collimator} 
%********************************************************************************************
Achieving high accuracy in laser beam collimation is challenging because the process is extremely sensitive 
to small mechanical and optical imperfections. For perfect collimation, the laser waist must be 
positioned exactly at the collimating lens’s focal point, and even micrometer level positional errors or microradian level angular misalignments introduce noticeable beam divergence error at large distance. 
Additionally, practical issues such as lens aberrations, surface irregularities, thermal drifts, 
and mechanical vibrations further degrade collimation quality. 
Real laser beams also deviate from an ideal Gaussian $M^{2} > 1$, causing 
asymmetric divergence that complicates precise adjustment. 
These challenges become even more significant for large diameter beams, where 
small errors are magnified over long propagation distances, making stable 
and high precision collimation difficult to achieve.
To address these challenges, we designed and assembled a beam collimator using commercially 
available off-the-shelf components that provides high precision kinematic alignment 
for accurate beam collimation.
\begin{figure}[t]
\includegraphics[width=0.49\textwidth]{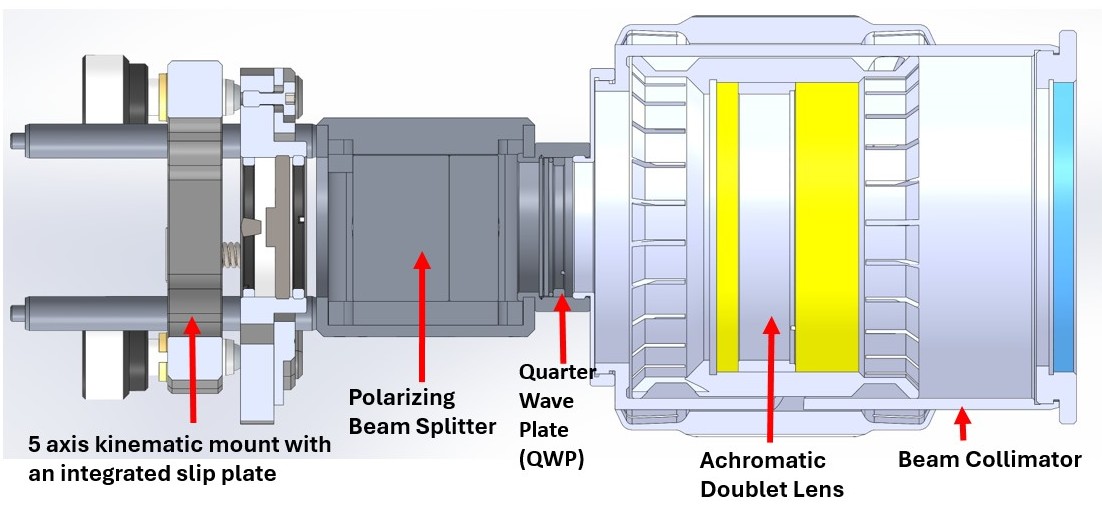}
\caption{\label{fig93} Three-dimensional mechanical model of the complete beam collimator. 
The overall length of the beam collimator is 250 mm.}
\end{figure}

The beam collimator is specifically designed for the cold-atom Gravimeter system 
currently under development at TCG CREST, Kolkata.
A three dimensional mechanical model of the complete beam collimator is presented in figure \ref{fig93}.
The beam collimator has a compact design, as the total length of the system is 250 mm. 
At the backend of the collimator, a 30 mm cage compatible threaded kinematic mount 
with an integrated slip plate is used. The slip plate holds threaded FC/APC fiber terminator adapters 
with a 2.0 mm narrow key slot, ensuring precise and repeatable coupling with the input fiber. 
This adapter effectively serves as the laser source for the collimator and enabling stable, 
well aligned beam injection into the system. 
The slip plate allows the laser source to move ±1 mm in the X and Y directions 
relative to the front plate, providing additional flexibility when positioning the source along the 
optical axis of the cage system.
Once the desired position is achieved, three locking screws allow the slip plate to be securely fixed in place.
Furthermore, the kinematic mount includes three precision adjusters that offer 0.4° (7 mrad) of angular adjustment 
per revolution and support up to ±4° of tip and tilt, along with ±3 mm of linear translation along the optical axis. 
Each adjuster can be independently locked using a conveniently placed 2.0 mm hex setscrew, ensuring 
stable and reliable alignment of the laser source.
Such high precision adjusters are essential for achieving a high degree of beam collimation and 
controlling the beam quality by positioning the laser source at the correct location and angle
such that its optical axis is referenced to the optical axis of the lens. 

\begin{figure}
\includegraphics[width=0.48\textwidth]{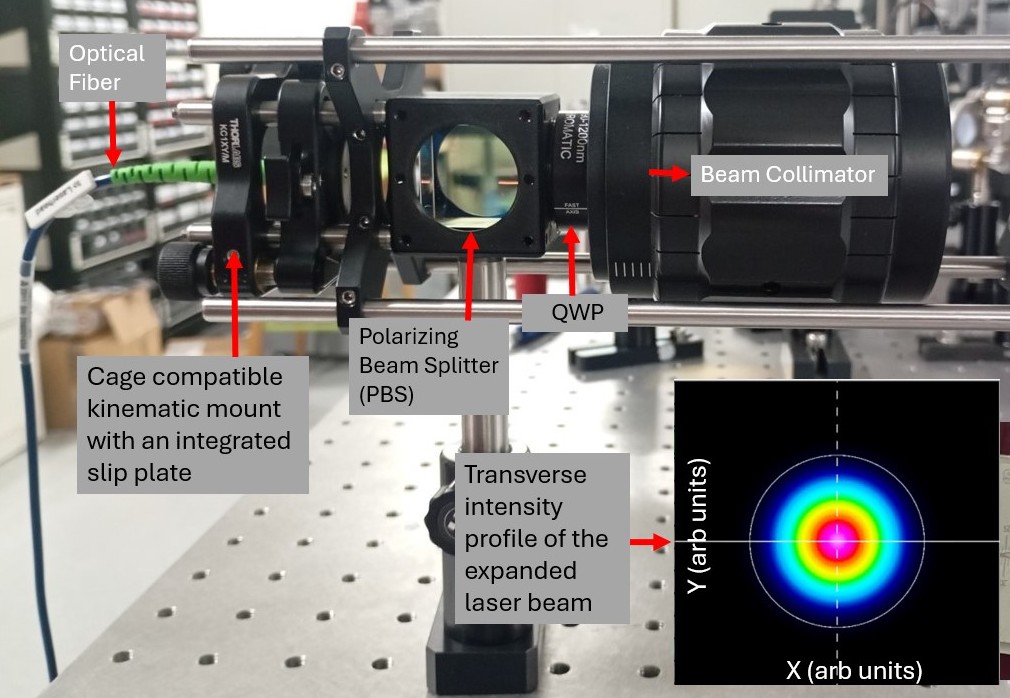}
\caption{\label{fig2:epsart} (a) Optical design of the experimental beam expander setup. 
(b) Transverse intensity profile of the expanded laser beam. (c) Normalised intensity profile of the laser beam.}
\end{figure}

At the front end of the beam collimator, a high precision adjustable collimation adapter is installed. 
This adapter securely holds a removable 100 mm AR coated achromatic doublet collimation lens. 
The lens is mounted on an internal carriage that provides a non-telescoping, rotational translation 
along the z-axis. This translation is achieved by turning the knurled adjustment ring.
The ring position is locked by turning the locking screw on the side of the adjustment ring with a 2 mm hex key.
The adapter allows up to 20 mm of linear travel for precise control over the lens position. 
The threading on the input and output apertures of the adapter remain fixed during translation.
The collimation adapter can be rotated clockwise or counterclockwise with extremely high precision. 
This rotational adjustment provides fine control over the beam deviation angle, ensuring 
optimal alignment and collimation performance.
The beam collimator produces an exceptionally well collimated laser beam with 
a diameter of 16 mm. A photograph of the assembled beam collimator is presented in figure \ref{fig2:epsart}. 
Its compact design, high alignment precision, and cost effectiveness make it well suited 
for cold-atom Gravimeter system and other precision optical applications.

Note that a polarizing beam splitter (PBS) and a quarter-wave plate (QWP) are installed 
at the center of the beam collimator. These two optical elements are included solely for 
the cold-atom Gravimeter system and do not influence the beam quality measurement or the degree of collimation.

%********************************************************************************************
\section{\label{sec3} Beam Quality and Collimation Characterization} 
%********************************************************************************************
Several techniques are available for assessing optical beam collimation, most of which are based on 
shearing interferometry or self imaging methods \cite{5000446,Mudassar:12,Herrera-Fernandez_2016,Rana:18}. 
Among these, Murty's plane parallel plate shearing interferometer is the simplest approach, in 
which a well collimated beam produces a fringe free overlap region between the sheared 
beams \cite{Murty:64}.
For incoherent beams, diffraction grating-based techniques have been introduced to enable 
reliable collimation measurements \cite{Torcal-Milla:17,Shakher:01,Sanchez-Brea:10,1.5000446,Anand:05}. 
In this section, we present a simple and efficient experimental procedure for 
determining the beam quality factor and quantitatively evaluating the degree of 
collimation of the laser beam produced by the developed beam collimator.

The $M^{2}$ value is an important indicator of beam quality and its determination requires careful beam
propagation measurements along with a precise measurement of the beam waist diameter ($d_{0}$).
For this purpose, the laser beam under test is passed through a plano convex AR coated 
lens with a focal length of $f = 250~mm$.
After passing through the lens, the laser beam diameter decreases as it approaches the focal point. 
The smallest achievable beam waist diameter $d_{0}$ is determined by the diffraction limit, which 
depends on the wavelength and the beam's divergence.
The beam cannot be focused below this fundamental limit. 
Beyond the waist, the beam diameter increases again during propagation. 
The beam quality factor $M^{2}$ characterizes how close the diffraction limit 
of the analyzed beam is to the diffraction limit of an ideal Gaussian beam with $M^{2} = 1$.
For this purpose, a beam propagation measurement is performed in accordance with 
the ISO11146 standard \cite{10.10631.2198795,Richter:07,MONTERO2024101830}.
\begin{figure}[t]
\begin{center}
  \includegraphics[scale=0.2]{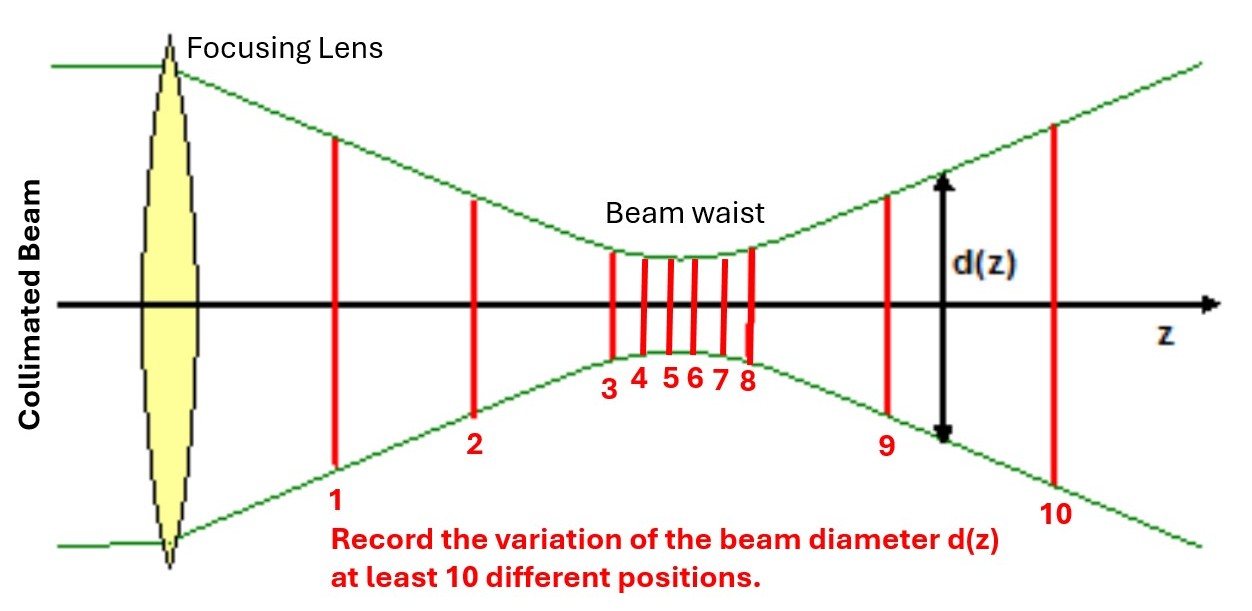}
  \caption{Schematic diagram of the experimental setup used to measure the beam waist, divergence angle, 
           Rayleigh length, and the beam quality factor $M^2$. 
           Collimated laser beam passes through a focusing lens of focal length, $f=250~mm$ and beam diameter 
           decreases as it approaches the len's focal point.}
    \label{panel}
\end{center}
\end{figure}
The schematic diagram of the measurement setup used in this work is presented in figure \ref{panel}.
The key idea is to record the variation of the beam diameter $d(z)$ at least 10 different positions along 
the propagation axis $z$.
The beam diameters along the $x$- and $y$-axes, denoted as $d_{x}(z)$ and $d_{y}(z)$, are 
measured using a commercial beam profiler. 
The propagation distance is varied using a motorized translation stage 
to record the beam diameter at different positions. 
As illustrated in figure \ref{panel}, a higher density of measurement points 
are taken near the beam waist to ensure an accurate determination of the minimum beam diameter, $d_0$.
The beam divergence angle, $\theta$ is a critically important parameter because it determines
how rapidly a laser beam spreads as it propagates away from the waist.
The beam diameter is related to the propagation distance through the following expression:
\begin{equation}\label{Eq:1}
d(z)=d_{0}\sqrt{1+\left(\frac{z}{z_{R}}\right) ^{2}}
\end{equation}
From the expression of $d(z)$, it is evident that in the far field $z >> z_R$, the beam diameter increases 
linearly with the propagation distance $z$.
The full divergence angle, $\theta$ can be determined as
\begin{equation}\label{Eq:2}
\theta \sim\frac{d}{z}=\frac{d_{0}}{z_{R}}
\end{equation}
As illustrated in figure \ref{panel}, the beam waist and divergence angle are accurately determined 
by analyzing the measured variation of the beam diameter along the propagation direction, $z$.
Once these parameters are obtained, the beam quality factor $M^{2}$ can be calculated using the following equation
\begin{equation}\label{Eq:6}
M^{2}=\frac{\pi}{4\lambda} d_0\times\theta   
\end{equation}
Equation \ref{Eq:6} clearly shows that accurate determination of the laser beam quality 
factor, $M^{2}$, fundamentally depends on precise measurements of the beam waist and divergence angle. 
Any uncertainty in these parameters propagates through the subsequent calculations, resulting 
in errors in the estimated $M^{2}$ value. 
Consequently, accurate characterization of the beam waist and divergence angle constitute 
the essential first step toward obtaining a reliable evaluation of the beam quality factor.

\begin{figure}[t]
\begin{center}
  \includegraphics[scale=0.19]{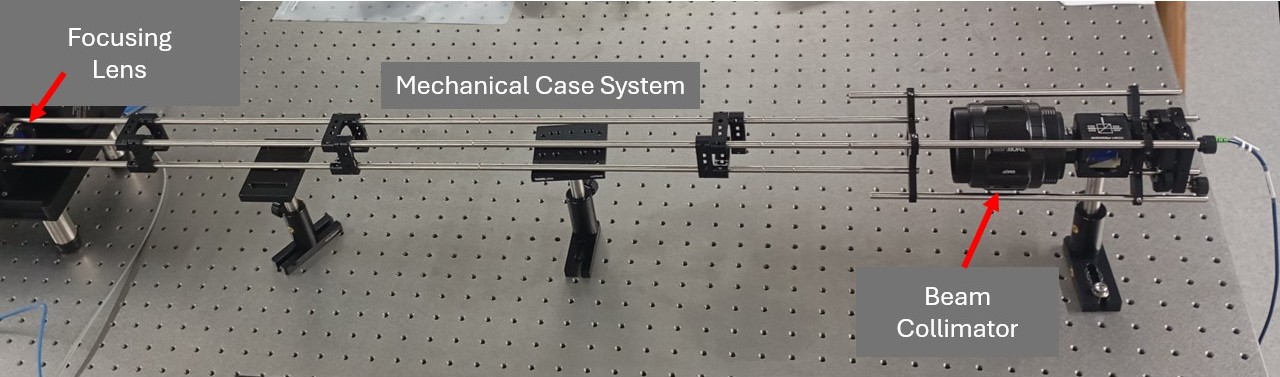}
  \caption{A tabletop experimental setup is used to achieve high precision beam alignment. 
  A high precision opto-mechanical caging and alignment platform system with a length of 1050 mm is integrated 
  with the beam collimator unit.}
  \label{Expsetup}
\end{center}
\end{figure}
%********************************************************************************************    
\section{\label{sec4} Methodology for High Precision Beam Alignment} 
%********************************************************************************************
Precise alignment of the test laser beam with the optical axis of the $M^{2}$ measurement setup 
and the beam profiler is essential for obtaining accurate results. 
The beam center must pass exactly through the center of the beam profiler's 9 mm aperture 
at every position of the translation stage. 
Even a few micrometers of linear translation or a few microradians of angular deviation in either the 
$x$- or $y$-direction can shift the beam off-center, causing it to clip or completely miss the aperture.
To address this challenge especially given that the beam collimator is located approximately 1050 mm from 
the plano-convex lens of the $M^{2}$ measurement system. 
We have designed and implemented a high precision opto-mechanical caging and alignment platform
as shown in the tabletop experimental setup in figure \ref{Expsetup}. 
This system enables fine control of the beam position and beam angle 
with micrometer level precision, ensuring stable and repeatable beam alignment throughout the measurement.

The beam alignment process is carried out in two sequential stages: coarse alignment and fine alignment.
\begin{figure}[t]
\begin{center}
  \includegraphics[scale=0.32]{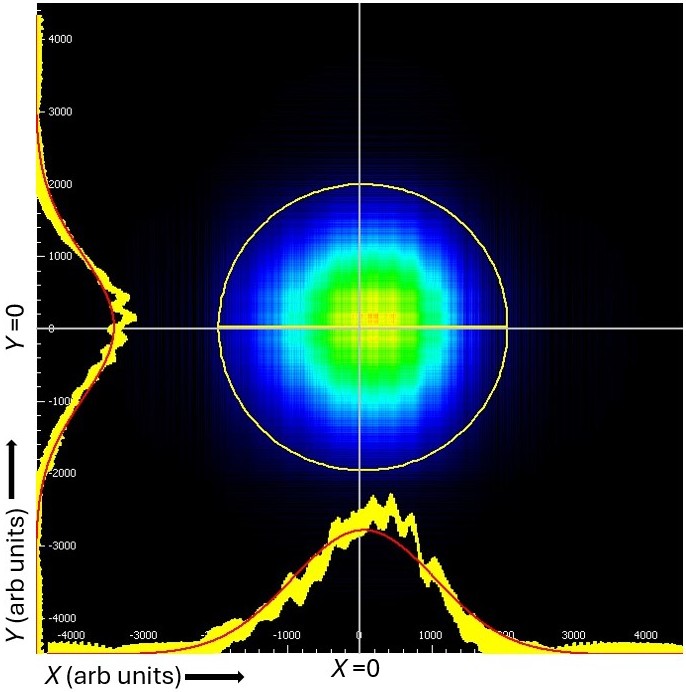}
  \caption{The coarse alignment procedure requires adjusting the beam so that laser beam profile is positioned as close as 
           possible to the center of the beam profiler's crosshair.
           The yellow curves in the horizontal and vertical directions represent the intensity profiles along 
           the $x$- and $y$-axes, respectively.}
    \label{position1}
\end{center}
\end{figure}

a. Coarse Alignment:
In the first step, the overall position of the laser source is adjusted to ensure that the beam follows 
the approximate optical path of the $M^{2}$ measurement system. 
Using the laser's built in control knobs, the beam is guided toward the intended trajectory. 
With the aid of the beam profiler, the beam is then aligned as close as possible to the center 
of the on-screen crosshair ($X=0, Y=0$), as illustrated in figure \ref{position1}.
This step is highly sensitive; even slight adjustments can cause noticeable shifts in the beam position. 
Therefore, achieving proper coarse alignment requires patience and meticulous control.

b. Fine Alignment:
After the beam is brought near the desired path, precise optimization is performed using 
the alignment tools provided in the mounting knobs that holds the plano convex lens. 
This fine tuning step corrects residual positional and angular deviations, ensuring 
that the beam is accurately centered and properly pointed at every stage position.
As shown in figure \ref{panel2}, when the fine alignment is optimized, the peak of the beam 
consistently remains within the central square
of the beam profiler display throughout the entire scan range.
As the beam alignment is progressively refined, the dimensions of the central square, defined by $X_2 - X_1$ 
and $Y_2 - Y_1$ decrease correspondingly.
Fine alignment is therefore crucial for maintaining consistent coupling through the $9~mm$ aperture
of the beam profiler and for obtaining reliable $M^{2}$ measurements.
\begin{figure}[t]
\begin{center}
  \includegraphics[scale=0.37]{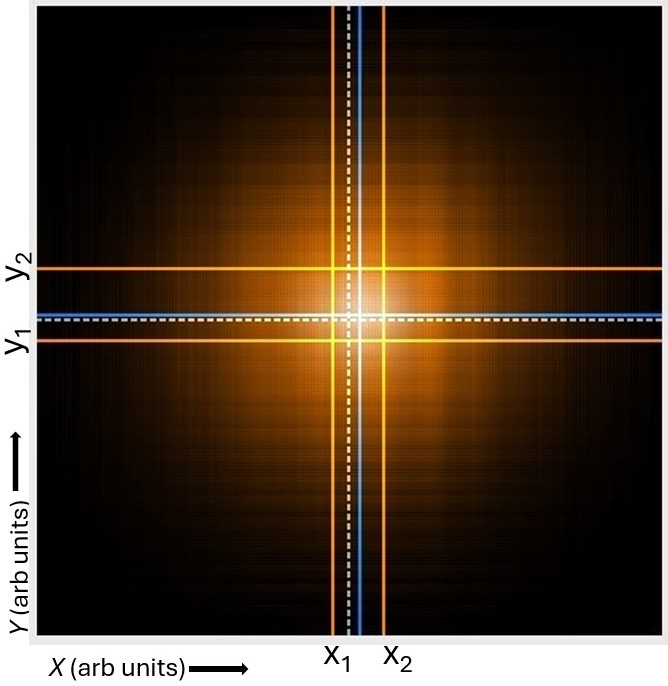}
  \caption{The fine tuning step corrects any remaining positional and angular deviations, ensuring 
           that the beam is precisely centered and properly aligned at every stage position. When the 
           alignment is optimized, the beam peak consistently remains within the central square 
           of the profiler display throughout the entire scan range.}
    \label{panel2}
\end{center}
\end{figure}

%********************************************************************************************
\section{\label{sec5} Results and Discussions}
%********************************************************************************************
Before beginning any $M^{2}$ measurement, we ensure that several essential conditions 
are met to guarantee reliable and accurate results.
First, the laser beam must be aligned correctly so that it remains centered 
on the detector across the entire scan range.
Additionally, the beam diameter must satisfy the specified requirements of the measurement 
system to avoid clipping or systematic errors. 
The exact wavelength of the laser under test must also be known, as even small 
uncertainties in wavelength can influence the calculated $M^{2}$ value.
The wavelength of the laser is $\lambda=780.241~nm$ which is measured using a wavemeter. 
Finally, the laser output should exhibit both spatial and temporal stability 
throughout the measurement to ensure consistent beam characteristics and reproducible results.

\begin{figure}[t]
\begin{center}
  \includegraphics[scale=0.18]{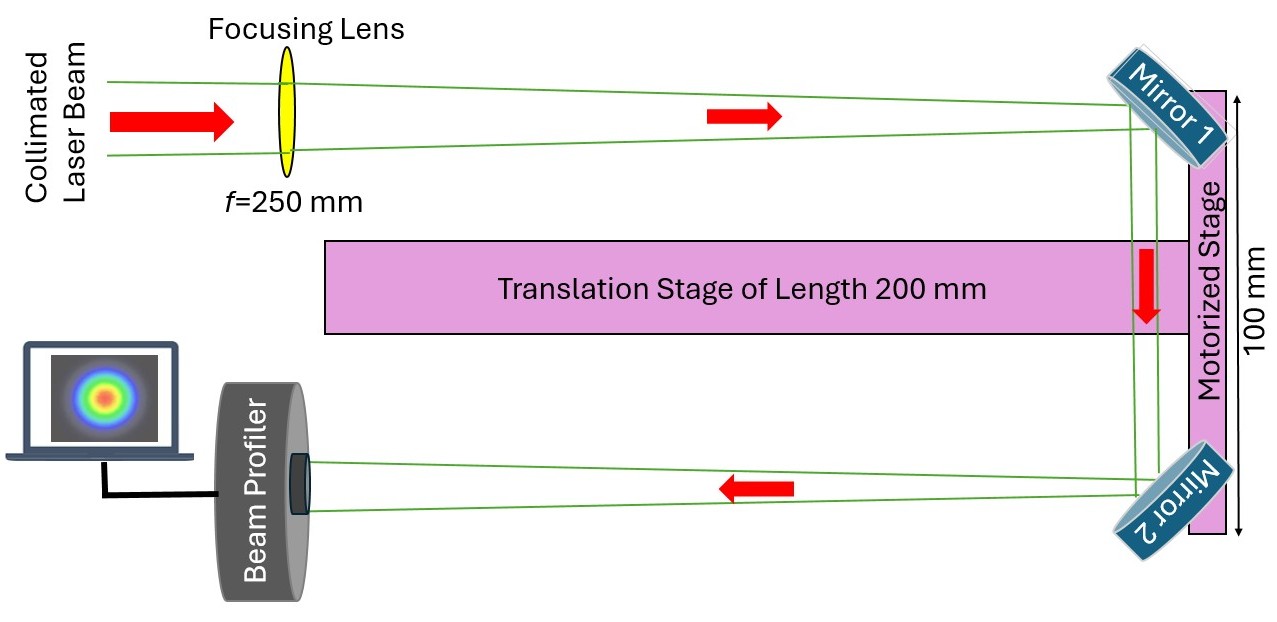}
  \caption{Schematic diagram of beam collimation testing setup. Two highly reflective mirrors, separated 
           by a fixed distance of $100~mm$, are mounted on a motorized stage. 
           The motorized stage translates along a linear translation stage with a travel range of $200~mm$.}
    \label{position25}
\end{center}
\end{figure}

A schematic of the beam collimation testing setup is shown in figure \ref{position25}. 
Two highly reflective mirrors, separated by a fixed distance of $100~mm$, are mounted on a motorized stage. 
This motorized stage translates along a linear translation stage with a travel range of 200 mm, 
measured from a focusing lens of focal length $250~mm$. 
Measurements are performed at various positions of the motorized stage as it is scanned from 0 to 200 mm.
The effective propagation distance of the collimated laser beam therefore varies from 100 mm to 500 mm, depending 
on the stage position. For instance, when the motorized stage is positioned at 200 mm, the total 
propagation distance, $z$ of the collimated beam is
\begin{equation}
z = 200~\text{mm} + 100~\text{mm} + 200~\text{mm} = 500~\text{mm}.
\end{equation}

\begin{figure}[t]
\begin{center}
  \includegraphics[scale=0.42]{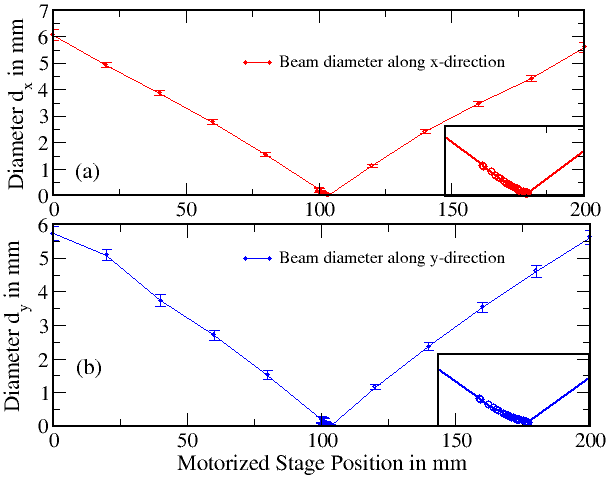}
  \caption{Beam diameter variations are plotted at various positions of the motorized stage 
           as it is scanned from 0 to 200 mm. 
           The red curve represents the beam diameter along the $x$-direction, while 
           the blue curve corresponds to the beam diameter along the $y$-direction.
           The insets depict a magnified representation of the beam diameter in the vicinity of the beam waist region.}
    \label{p7}
\end{center}
\end{figure}

We recorded the beam diameter at multiple positions of the motorized stage and 
the resulting variation is plotted in figure \ref{p7}.
Figure \ref{p7}(a) shows the beam diameter variation along the $x$-axis, while figure \ref{p7}(b) presents 
the corresponding variation along the $y$-axis.
As shown in figure \ref{p7}, we consider a higher density of measurement points near the beam waist 
to accurately determine the minimum beam diameter.
Now, from the curve fitting analysis of the measured data, we obtain the minimum beam diameters 
corresponding to the beam waist. 
The fitted beam waist along the $x$-axis is $d_{0x} = 23.03~ \mu\text{m}$, while the waist 
along the $y$-axis is $d_{0y} = 30.59~ \mu\text{m}$. 
These values represent the smallest beam diameters achieved in their respective 
directions and serve as key parameters for evaluating the overall beam quality.
We observe a clear asymmetry in the measured beam diameters along the $x$ and $y$ axes. 
To quantify this, the beam waist asymmetry (BWA) is defined as
\begin{equation}\label{Eq:3}
BWA=\frac{d_{0y}}{d_{0x}}
\end{equation}
where $d_{0x}$ and $d_{0y}$ are the minimum beam diameters along the respective axes. 
The calculated BWA value of 1.33 confirms a significant difference in the waist sizes 
between the two directions.
The fitted beam waist positions are 106.82 mm along the $x$ axis and 106.64 mm along the $y$ axis, 
indicating that the focal positions are essentially coincident despite the asymmetry in waist size. 
The corresponding Rayleigh lengths are 0.36 mm for the $x$ axis and 0.47 mm for the $y$ axis. 
This difference reflects the variation in beam divergence caused by the unequal waist diameters: the 
axis with the smaller waist exhibits a shorter Rayleigh length and therefore a larger divergence.

Next, we determine the beam divergence along each axis. 
The divergence angle along the $x$ axis is calculated using $\theta_x=\frac{d_{0x}}{z_{Rx}}$
which yields a value of $\theta_x = 3.654^\circ$ for the beam after passing through the lens. 
Following the same procedure for the $y$ axis, we obtain a divergence angle of $\theta_y = 3.738^\circ$.
From these values, the divergence asymmetry, defined as 
the ratio $\theta_y / \theta_x$, is found to be 1.023, indicating 
that the beam diverges slightly more along the $y$ axis than along the $x$ axis. 
This small but noticeable difference is consistent with the asymmetry observed in the beam waist measurements.

Next, we calculate the beam quality factor $M^{2}$, which provides a 
quantitative measure of how closely the analyzed beam approaches an ideal Gaussian beam. 
Determining $M^{2}$ requires two essential parameters: the beam waist, which specifies 
the minimum achievable spot size, and the divergence angle, which describes how quickly 
the beam expands during propagation. By combining these parameters, we obtain a 
complete description of the beam's deviation from the diffraction-limited case.
Based on the measured waist sizes and divergences, the resulting beam quality factors are $M^{2}_{x} = 1.48$ 
and $M^{2}_{y} = 2.01$. 
These values indicate moderate departures from an ideal Gaussian profile, with 
the beam exhibiting slightly better quality along the $x$ axis compared to the $y$ axis.

Figure \ref{panel} illustrates how a laser beam propagates through a thin lens, leading 
to both a shift in the beam waist position and a change in its size, along with corresponding modifications 
to the Rayleigh length $z_{R}$ and divergence angle $\theta$. 
When a collimated beam passes through a focusing lens, the new beam waist 
formed after the lens is determined primarily by the product of the lens focal 
length and the divergence of the incoming beam.
From this relationship, the divergence angle of the original (input) beam can be expressed as
\begin{equation}
\theta_{\text{org}} = \frac{d_{0}}{f},
\end{equation}
where $d_{0}$ is the measured beam waist after the lens and $f$ is the focal length of the lens.
Using this expression, the original divergence angles are calculated to be 0.006° (105 µrad) along the $x$
axis and 0.007° (122 µrad) along the $y$ axis. 
These exceptionally low divergence values confirm the excellent collimation of the laser beam.
This high degree of collimation makes the designed beam collimator well suited for cold-atom Gravimeter setup.

%********************************************************************************************
\section{\label{sec6} Implementation of the Designed Beam Collimator in the Cold-Atom Gravimeter system}
%********************************************************************************************
\begin{figure}[t]
\includegraphics[width=0.42\textwidth]{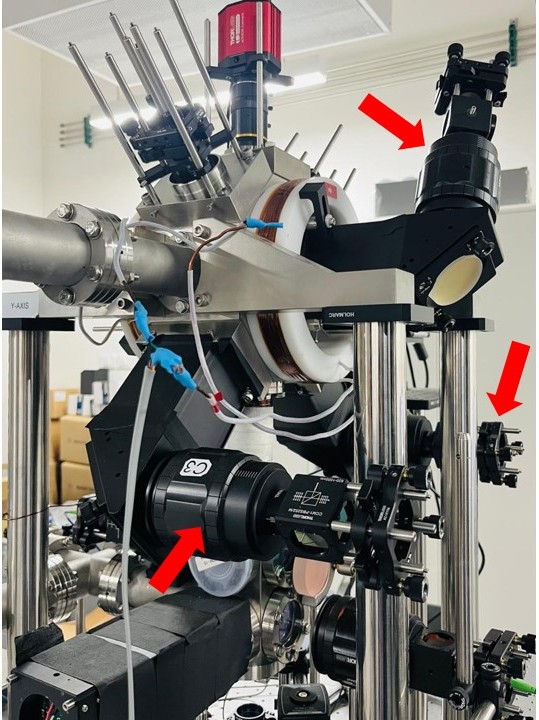}
\caption{\label{fig10} The high precision beam collimators mounted on the Magneto-Optical Trap chamber are highlighted with red arrows.
The collimators are mounted to ensure precise delivery of expanded and well-collimated laser beams 
required for efficient atom trapping and cooling.}
\end{figure}
The developed beam collimator has been successfully implemented in the 
cold-atom Gravimeter under development at TCG CREST, Kolkata. 
In figure \ref{fig10}, the red arrows show the integration of the designed beam collimator 
into the complete cold-atom Gravimeter system. 
The collimated laser beams generated by this device are used to form 
the six-beam $^{87}$Rb Magneto-Optical Trap, which currently demonstrates 
a loading time of approximately 210 ms and a trap lifetime of approximately 540 ms. 
These results confirm the suitability of the beam collimator for cold-atom experiments 
requiring stable and highly uniform laser beams.

Accurate beam collimation is a critical requirement for efficient laser cooling and trapping. 
Well-collimated laser beams ensure uniform intensity distribution and proper overlap 
at the trapping region, thereby maximizing the number of trapped atoms while minimizing wavefront distortions 
and beam divergence. Since the sensitivity and precision of a cold-atom quantum Gravimeter 
strongly depend on the number of atoms participating in the interferometer, efficient atom trapping 
directly contributes to improved measurements of the local acceleration due to gravity ($g$). 
The compact beam collimator presented in this work therefore provides a reliable, high-performance, 
and cost-effective solution for cold-atom applications. 
Its robust mechanical design, high-precision alignment capability, and excellent beam quality 
make it well suited not only for quantum Gravimeter but also for a wide range of laser cooling, atom interferometry, 
and other precision quantum sensing experiments.

%********************************************************************************************
\section{\label{sec7} Conclusion}
%********************************************************************************************
In this study, we present a precisely engineered laser beam collimation system 
and a systematic beam quality evaluation methodology, both developed to meet the stringent 
performance requirements of a cold-atom Gravimeter \cite{Anju_2025}.
Maintaining a highly collimated beam over long propagation distances is essential 
in quantum sensing applications, where small deviations in beam divergence 
can accumulate and degrade measurement accuracy.
By employing Gaussian beam theory and an ISO11146-based \cite{MONTERO2024101830} propagation analysis, we accurately 
extracted the beam waist, divergence angle, and beam quality factor \cite{10.10631.2198795,Richter:07}. 
The high precision beam collimator combines high resolution 5-axis kinematic control at the source 
with a finely adjustable achromatic doublet output lens, enabling independent optimization of beam position, 
angle, and longitudinal focus. This configuration allows microradian level collimation control 
and stable propagation of a 16 mm diameter beam.
The experimentally measured divergence angles, on the order of $100$ micro-radian, demonstrate 
effective long range collimation, while the measured $M^{2}$ values capture residual 
transverse asymmetry without limiting practical performance. The demonstrated approach 
offers a robust and adaptable solution for cold atom interferometry \cite{Anju_2025,Tino_2021} 
and other precision optical systems that require stable, high quality laser beams over extended distances.

\begin{acknowledgments}
N.S.M. acknowledge funding from the Anusandhan National Research Foundation (ANRF)
through the ANRF National Post-Doctoral Fellowship (Grant No. PDF/2025/002172).
Anju would like to thank Qmet Tech Foundation for postdoctoral fellowship.
All authors thank the TCG CREST, India, for providing the facilities to carry out this work.
\end{acknowledgments}

\nocite{*}
\bibliography{reference}% Produces the bibliography via BibTeX.

@CONTROL{REVTEX41Control}

@CONTROL{aip41Control,pages="1",title="0"}

@article{PhysRevResearch.6.L032028,
  title = {Accurate and efficient Bloch-oscillation-enhanced atom interferometry},
  author = {Fitzek, F. and Kirsten-Siem\ss{}, J.-N. and Rasel, E. M. and Gaaloul, N. and Hammerer, K.},
  journal = {Phys. Rev. Res.},
  volume = {6},
  issue = {3},
  pages = {L032028},
  numpages = {6},
  year = {2024},
  month = {Aug},
  publisher = {American Physical Society},
  doi = {10.1103/PhysRevResearch.6.L032028},
  url = {https://link.aps.org/doi/10.1103/PhysRevResearch.6.L032028}
}

@article{1.5000446,
    author = {Luo, Xiaohe and Hui, Mei and Wang, Shanshan and Hou, Yinlong and Zhou, Siyu and Zhu, Qiudong},
    title = {Collimation testing using slit Fresnel diffraction},
    journal = {Review of Scientific Instruments},
    volume = {89},
    number = {3},
    pages = {033102},
    year = {2018},
    month = {03},
    issn = {0034-6748},
    doi = {10.1063/1.5000446},
    url = {https://doi.org/10.1063/1.5000446},
}

@article{Anand:05,
author = {Arun Anand and Vani K. Chhaniwal and C. S. Narayanamurthy},
journal = {Appl. Opt.},
number = {20},
pages = {4244--4247},
publisher = {Optica Publishing Group},
title = {Collimation testing with optically active materials},
volume = {44},
month = {Jul},
year = {2005},
url = {https://opg.optica.org/ao/abstract.cfm?URI=ao-44-20-4244},
doi = {10.1364/AO.44.004244},
}

@article{Prakash10032006,
author = {Shashi Prakash and Santosh Rana},
title = {A modified approach for collimation testing using Lau interferometry},
journal = {Journal of Modern Optics},
volume = {53},
number = {4},
pages = {507--512},
year = {2006},
publisher = {Taylor \& Francis},
doi = {10.1080/09500340500420373},
URL = { https://doi.org/10.1080/09500340500420373},
}

@article{Rana:18,
author = {Santosh Rana and Jitendra Dhanotia and Vimal Bhatia and Shashi Prakash},
journal = {Appl. Opt.},
number = {10},
pages = {2686--2692},
publisher = {Optica Publishing Group},
title = {Automated collimation testing by determining the statistical correlation coefficient of Talbot self-images},
volume = {57},
month = {Apr},
year = {2018},
url = {https://opg.optica.org/ao/abstract.cfm?URI=ao-57-10-2686},
doi = {10.1364/AO.57.002686},
}

@article{Herrera-Fernandez_2016,
doi = {10.1088/2040-8978/18/7/075608},
url = {https://doi.org/10.1088/2040-8978/18/7/075608},
year = {2016},
month = {jun},
publisher = {IOP Publishing},
volume = {18},
number = {7},
pages = {075608},
author = {Herrera-Fernandez, Jose Maria and Sanchez-Brea, Luis Miguel and Torcal-Milla, Francisco Jose and Morlanes, Tomas and Bernabeu, Eusebio},
title = {Dual self-image technique for beam collimation},
journal = {Journal of Optics}
}

@article{Sanchez-Brea:10,
author = {Luis Miguel Sanchez-Brea and Francisco Jose Torcal-Milla and Francisco Javier Salgado-Remacha and Tomas Morlanes and Isidoro Jimenez-Castillo and Eusebio Bernabeu},
journal = {Appl. Opt.},
number = {17},
pages = {3363--3368},
publisher = {Optica Publishing Group},
title = {Collimation method using a double grating system},
volume = {49},
month = {Jun},
year = {2010},
url = {https://opg.optica.org/ao/abstract.cfm?URI=ao-49-17-3363},
doi = {10.1364/AO.49.003363},
}

@article{Shakher:01,
author = {Chandra Shakher and Shashi Prakash and Daya Nand and Rajesh Kumar},
journal = {Appl. Opt.},
number = {8},
pages = {1175--1179},
publisher = {Optica Publishing Group},
title = {Collimation testing with circular gratings},
volume = {40},
month = {Mar},
year = {2001},
url = {https://opg.optica.org/ao/abstract.cfm?URI=ao-40-8-1175},
doi = {10.1364/AO.40.001175},
}

@article{Mudassar:12,
author = {Asloob Ahmad Mudassar and Saira Butt},
journal = {Appl. Opt.},
number = {26},
pages = {6429--6440},
publisher = {Optica Publishing Group},
title = {Improved collimation testing technique},
volume = {51},
month = {Sep},
year = {2012},
url = {https://opg.optica.org/ao/abstract.cfm?URI=ao-51-26-6429},
doi = {10.1364/AO.51.006429},
}

@Article{photonics13020142,
AUTHOR = {Liu, Fan and Zhang, Hui and Bai, Yang and Ruan, Jun and Yang, Shaojie and Zhang, Shougang},
TITLE = {Development of a Compact Laser Collimating and Beam-Expanding Telescope for an Integrated 87Rb Atomic Fountain Clock},
JOURNAL = {Photonics},
VOLUME = {13},
YEAR = {2026},
NUMBER = {2},
ARTICLE-NUMBER = {142},
URL = {https://www.mdpi.com/2304-6732/13/2/142},
ISSN = {2304-6732},
DOI = {10.3390/photonics13020142}
}

@article{10.10635.0198240,
    author = {Chen, Hong-Hui and Yao, Zhan-Wei and Lu, Ze-Xi and Lu, Si-Bin and Jiang, Min and Li, Shao-Kang and Chen, Xiao-Li and Sun, Chuan and Mao, Yin-Fei and Li, Yang and Li, Run-Bing and Wang, Jin and Zhan, Ming-Sheng},
    title = {Self-calibrated atom-interferometer gyroscope by modulating atomic velocities},
    journal = {Review of Scientific Instruments},
    volume = {95},
    number = {5},
    pages = {053201},
    year = {2024},
    month = {05},
    issn = {0034-6748},
    doi = {10.1063/5.0198240},
    url = {https://doi.org/10.1063/5.0198240},
}

@article{MONTERO2024101830,
title = {Beamdiameter — Development of software for a laser beam profiler},
journal = {SoftwareX},
volume = {27},
pages = {101830},
year = {2024},
issn = {2352-7110},
doi = {https://doi.org/10.1016/j.softx.2024.101830},
url = {https://www.sciencedirect.com/science/article/pii/S2352711024002012},
author = {Mario Ricardo Montero and Juan Carlos Alvarez and Francisco Juan Racedo N.},
}

@article{Richter:07,
author = {Dirk Richter and Petter Weibring and Alan Fried and Osamu Tadanaga and Yoshiki Nishida and Masaki Asobe and Hiroyuki Suzuki},
journal = {Opt. Express},
number = {2},
pages = {564--571},
publisher = {Optica Publishing Group},
title = {High-power, tunable difference frequency generation source for absorption spectroscopy based on a ridge waveguide periodically poled lithium niobate crystal},
volume = {15},
month = {Jan},
year = {2007},
url = {https://opg.optica.org/oe/abstract.cfm?URI=oe-15-2-564},
doi = {10.1364/OE.15.000564},
}

@article{10.10631.2198795,
    author = {Schäfer, B. and Lübbecke, M. and Mann, K.},
    title = {Hartmann-Shack wave front measurements for real time determination of laser beam propagation parameters},
    journal = {Review of Scientific Instruments},
    volume = {77},
    number = {5},
    pages = {053103},
    year = {2006},
    month = {05},
    issn = {0034-6748},
    doi = {10.1063/1.2198795},
    url = {https://doi.org/10.1063/1.2198795},
}

@Article{atoms9030058,
AUTHOR = {Gochnauer, Daniel and Rahman, Tahiyat and Wirth-Singh, Anna and Gupta, Subhadeep},
TITLE = {Interferometry in an Atomic Fountain with Ytterbium Bose–Einstein Condensates},
JOURNAL = {Atoms},
VOLUME = {9},
YEAR = {2021},
NUMBER = {3},
ARTICLE-NUMBER = {58},
URL = {https://www.mdpi.com/2218-2004/9/3/58},
ISSN = {2218-2004},
DOI = {10.3390/atoms9030058}
}

@article{PhysRevApplied.23.044001,
  title = {Closed-loop measurements in an atom-interferometer gyroscope with compensation for velocity-dependent phase dispersion},
  author = {Sato, Tomoya and Nishimura, Naoki and Kaku, Naoki and Otabe, Sotatsu and Kawasaki, Takuya and Hosoya, Toshiyuki and Kozuma, Mikio},
  journal = {Phys. Rev. Appl.},
  volume = {23},
  issue = {4},
  pages = {044001},
  numpages = {10},
  year = {2025},
  month = {Apr},
  publisher = {American Physical Society},
  doi = {10.1103/PhysRevApplied.23.044001},
  url = {https://link.aps.org/doi/10.1103/PhysRevApplied.23.044001}
}

@article{Gustavson_2000,
doi = {10.1088/0264-9381/17/12/311},
url = {https://doi.org/10.1088/0264-9381/17/12/311},
year = {2000},
month = {jun},
publisher = {},
volume = {17},
number = {12},
pages = {2385},
author = {T L Gustavson and A Landragin and M A Kasevich},
title = {Rotation sensing with a dual atom-interferometer Sagnac gyroscope},
journal = {Classical and Quantum Gravity},
}

@article{Anju_2025,
doi = {10.1088/1402-4896/ae1c78},
url = {https://doi.org/10.1088/1402-4896/ae1c78},
year = {2025},
month = {nov},
publisher = {IOP Publishing},
volume = {100},
number = {11},
pages = {115405},
author = {Anju and Rathore, Harish and Mallick, Nawaz Sarif and Bhardwaj, Abhishek and Acharya, Aishik},
title = {Development and characterization of 87Rb atomic ensemble for absolute quantum gravimeter},
journal = {Physica Scripta}
}

@article{5000446,
    author = {Luo, Xiaohe and Hui, Mei and Wang, Shanshan and Hou, Yinlong and Zhou, Siyu and Zhu, Qiudong},
    title = {Collimation testing using slit Fresnel diffraction},
    journal = {Review of Scientific Instruments},
    volume = {89},
    number = {3},
    pages = {033102},
    year = {2018},
    month = {03},
    issn = {0034-6748},
    doi = {10.1063/1.5000446},
    url = {https://doi.org/10.1063/1.5000446},
}

@article{801756,
    author = {Bidel, Yannick and Carraz, Olivier and Charrière, Renée and Cadoret, Malo and Zahzam, Nassim and Bresson, Alexandre},
    title = {Compact cold atom gravimeter for field applications},
    journal = {Applied Physics Letters},
    volume = {102},
    number = {14},
    pages = {144107},
    year = {2013},
    month = {04},
    issn = {0003-6951},
    doi = {10.1063/1.4801756},
    url = {https://doi.org/10.1063/1.4801756},
}

@article{Tino_2021,
doi = {10.1088/2058-9565/abd83e},
url = {https://doi.org/10.1088/2058-9565/abd83e},
year = {2021},
month = {mar},
publisher = {IOP Publishing},
volume = {6},
number = {2},
pages = {024014},
author = {Tino, Guglielmo M},
title = {Testing gravity with cold atom interferometry: results and prospects},
journal = {Quantum Science and Technology}
}

@article{Sanchez-Brea:14,
author = {Luis Miguel Sanchez-Brea and Francisco Jose Torcal-Milla and Jose Maria Herrera-Fernandez and Tomas Morlanes and Eusebio Bernabeu},
journal = {Opt. Lett.},
number = {19},
pages = {5764--5767},
publisher = {Optica Publishing Group},
title = {Self-imaging technique for beam collimation},
volume = {39},
month = {Oct},
year = {2014},
url = {https://opg.optica.org/ol/abstract.cfm?URI=ol-39-19-5764},
doi = {10.1364/OL.39.005764},
}

@article{Torcal-Milla:17,
author = {Francisco Jose Torcal-Milla and Luis Miguel Sanchez-Brea},
journal = {Appl. Opt.},
number = {12},
pages = {3628--3633},
publisher = {Optica Publishing Group},
title = {Collimation technique and testing applied to finite size polychromatic sources},
volume = {56},
month = {Apr},
year = {2017},
url = {https://opg.optica.org/ao/abstract.cfm?URI=ao-56-12-3628},
doi = {10.1364/AO.56.003628},
}

@article{Murty:64,
author = {M. V. R. K. Murty},
journal = {Appl. Opt.},
number = {4},
pages = {531--534},
publisher = {Optica Publishing Group},
title = {The Use of a Single Plane Parallel Plate as a Lateral Shearing Interferometer with a Visible Gas Laser Source},
volume = {3},
month = {Apr},
year = {1964},
url = {https://opg.optica.org/ao/abstract.cfm?URI=ao-3-4-531},
doi = {10.1364/AO.3.000531},
}

\end{document}